\documentstyle[psfig,draft]{mn}
\def\gs{\mathrel{\raise0.35ex\hbox{$\scriptstyle >$}\kern-0.6em 
\lower0.40ex\hbox{{$\scriptstyle \sim$}}}}
\def\ls{\mathrel{\raise0.35ex\hbox{$\scriptstyle <$}\kern-0.6em 
\lower0.40ex\hbox{{$\scriptstyle \sim$}}}}
\date{Received 1999 -- ---; accepted: 1999 -- --}

\begin{document}

\title{The Discovery of ERO Counterparts to Faint Submm Galaxies}

\author[Smail et al.]{Ian Smail,$^{\! 1}$ R.\,J.\ Ivison,$^{\! 2}$
J.-P.\ Kneib,$^{\! 3}$ L.\,L.\ Cowie,$^{\! 4}$
A.\,W.\ Blain,$^{\! 5}$ A.\,J.\ Barger,$^{\! 4}$\and
F.\,N.\ Owen$^{6}$ \& G.\ Morrison$^{6}$\\
\hbox{~}\\
$^1$ Department of Physics, University of Durham, South Road, 
Durham DH1 3LE, UK\\
$^2$ Department of Physics and Astronomy, University College
London, Gower Street, London WC1E 6BT, UK\\
$^3$ Observatoire Midi-Pyr\'en\'ees, CNRS-UMR5572,
14 Avenue E.\ Belin, 31400 Toulouse, France\\
$^4$ Institute of Astronomy, University of Hawaii, 2680
Woodlawn Drive, Honolulu, Hawaii HI\,96822, USA\\
$^5$ Cavendish Laboratory, Madingley Road, Cambridge
CB3 OHE, UK\\
$^6$ NRAO, P.O.\ Box 0, 1003 Lopezville Road, Socorro, NM 87801}

\maketitle

\begin{abstract}
We have used deep ground-based imaging in the near-infrared to search
for counterparts to the luminous submillimeter (submm) sources in the
catalog of Smail et al.\ (1998).  For the majority of the submm sources
the near-IR imaging supports the counterparts originally selected from
deep optical images. However, in two cases (10\% of the sample) we find
a relatively bright near-IR source close to the submm position, sources
that were unidentified in the deep {\it Hubble Space Telescope (HST)}
and ground-based $R$-band images used in Smail et al.\ (1998).  We
place limits on colours of these sources from deep high-resolution
Keck\,II imaging and find they have 2-$\sigma$ limits of $(I-K)\gs 6.8$
and $(I-K)\gs 6.0$ respectively. Both sources thus class as extremely
red objects (EROs).  Using the spectral properties of the submm source
in the radio and submm we argue that these EROs are probably the source
of the submm emission, rather than the bright spiral galaxies
previously identified by Smail et al.\ (1998).  This connection
provides important insights into the nature of the enigmatic ERO
population and faint submm galaxies in general. ~From the estimated
surface density of these submm-bright EROs we suggest that this class
accounts for the majority of the reddest members of the ERO population,
in good agreement with the preliminary conclusions of pointed submm
observations of individual EROs.  We conclude that the most extreme
EROs represent a population of dusty, ultraluminous galaxies at high
redshifts; further study of these will provide useful insights into the
nature of star formation in obscured galaxies in the early Universe.
The identification of similar counterparts in blank field submm surveys
will be extremely difficult owing to their faintness ($K\sim 20.5$,
$I\gs 26.5$).  Finally, we discuss the radio and submm properties of the two
submm-bright EROs discovered here and suggest that both galaxies lie at
$z\gs 2$.
\end{abstract}


\section{Introduction}

The nature of the population of extremely red objects (dubbed `EROs',
Elston et al.\ 1988; Hu \& Ridway 1994; Barger et al.\ 1999a; Cowie et
al.\ 1999) uncovered in deep near-IR surveys remains elusive.  Their
extreme colours, $(R-K)\geq 6$ or $(I - K) \gs 5$, could be produced by
a number of diverse factors: an evolved stellar population in a
high-redshift galaxy; a substantial contribution to the $K$-band flux
from a strong emission line; a very high-redshift galaxy ($z>6$--8)
with strong absorption from the Ly\,$\alpha$ forest shortward of the
$K$ band, or a dust-obscured, highly reddened galaxy.  These
explanations suggest that a wide range of galaxies may belong to this class.

The surface density of the ERO population is $\ls 0.01$/arcmin$^{-2}$
at $K\leq 19$ in blank fields, roughly similar to that of QSOs (Hu \& Ridgway
1994). The exact surface density of sources depends upon the passbands
and colours used to define an ERO as well as the magnitude limit
of the sample, reaching $\sim 0.1$ sources per square arcmin redder
than $(I-K)>5$ at $K=21$ (Cowie et al.\ 1999).  EROs also appear to
have higher surface densities, $\times 10$--100, in the vicinity of
high-redshift active galactic nuclei (AGN) compared to the general
field (Aragon-Salamanca et al.\ 1994; Cowie et al.\ 1994; Dey et
al.\ 1995; Elston et al.\ 1988; Graham et al.\ 1994; Graham \& Dey
1996; Hu \& Ridgway 1994; Yamada et al.\ 1997; Soifer et al.\ 1992).
This apparent clustering has led to speculation that at least some of
the EROs are physically associated with luminous AGN.  Such clustering
behaviour is expected within a hierarchical framework where we would
find other massive galaxies (detectable EROs at the AGN redshifts would
have rest-frame optical luminosities of 5--20\,$L^*$) forming in the
vicinity of the high-density peak signposted by the AGN.  Thus it is
possible that a substantial fraction of the EROs comprise a population
of luminous galaxies in the distant Universe and moreover one which is
highly biased.  Studying this population and testing this hypothesis
would provide much-needed information about the formation of galaxies
in high-density regions at high redshift and their subsequent evolution.

The heterogeneous nature of the ERO class has led to slow progress in
defining the properties of this population.  In particular, the
feature which defines EROs -- their extreme faintness in the optical
($I > 24$--25) compared to the near-IR -- means that for most of them
their redshifts and spectral properties are unknown.  However, as observations
in more wavebands are becoming available we are gradually building up
a picture of the nature of this population.  Working in this manner,
Cimatti et al.\ (1998) and Dey et al.\ (1999) have presented new
optical, near-IR and submm observations of ERO\,J164502+4626.4 (also
known as HR\,10, an ERO discovered in the field of the binary quasar
PC\,1643+4631; Hu \& Ridgway 1994) which demonstrate convincingly that
it is a dusty, ultraluminous starburst galaxy at $z=1.44$.  The
interpretation of HR\,10 as a dusty starburst is based upon photometry
from the SCUBA bolometer array which shows that the far-IR luminosity
of HR\,10 is immense --- its bolometric luminosity of $7 \times
10^{12}$~L$_{\odot}$ places HR\,10 into the ultraluminous IR galaxy
class (ULIRG) and implies that stars are being formed at a rate of
1000--2000 M$_{\odot}$~yr$^{-1}$.  This is consistent with the
disturbed morphology of HR\,10: {\it HST} imaging apparently shows a
complex interaction or an ongoing merger, very similar to those that
trigger local ULIRGs (Sanders \& Mirabel 1996) as well as their more
distant counterparts (Smail et al.\ 1998).

The role of dust in defining the visible properties of EROs has been
tested using SCUBA, with observations of around twenty EROs now
completed at 850\,$\mu$m (Cimatti et al.\ 1998; Dey et al.\ 1999;
Andreani et al.\ 1999; Thommes et al.\ 1999). The picture that is
emerging is of a bimodal population --- roughly two thirds of the 
$K\ls 20$ EROs
are undetected in the submm, indicating that they are neither strongly
star forming nor contain large quantities of cold dust and thus are
probably red because of old stellar populations. The remainder, which
are typically redder, do show submm emission and have extreme colours
because of reddening by dust (these include HR\,10).  The surface
density of these dust-rich EROs, while highly uncertain, is such that
they could be a moderate fraction of the SCUBA sources brighter than a
few mJy at 850\,$\mu$m.  The confirmation of such a connection between
EROs and the SCUBA sources would provide important insights into the
nature of both of these populations.

In this paper we present multi-wavelength observations of two luminous
submm sources selected from the SCUBA Lens Survey of Smail et
al.\ (1998, S98).  These sources were initially identified with bright
spiral galaxies at $z=0.18$ and $z=0.33$.  Subsequent near-IR imaging
uncovered relatively bright galaxies close to the submm positions,
sources that were invisible on the original deep optical images.
Sensitive radio mapping with the Very Large Array (VLA) has
strengthened the identification of one of these red galaxies with the
submm emission.  We begin in \S2 by presenting the near-IR and radio
observations of these fields, along with new deep high-resolution
optical imaging to provide more stringent and uniform limits on the
colours of the galaxies.  We then discuss in \S3 the spectral energy
distributions (SEDs) of these two galaxies and compare these with what
is known of other EROs.  We give our conclusions in \S4.  Throughout
this paper we adopt $H_\circ = 50$\,km\,s$^{-1}$\,Mpc$^{-1}$ and
a $\Omega_\circ=1$, $\Lambda=0$ cosmogony.

%
%
\begin{figure*}
\centerline{\psfig{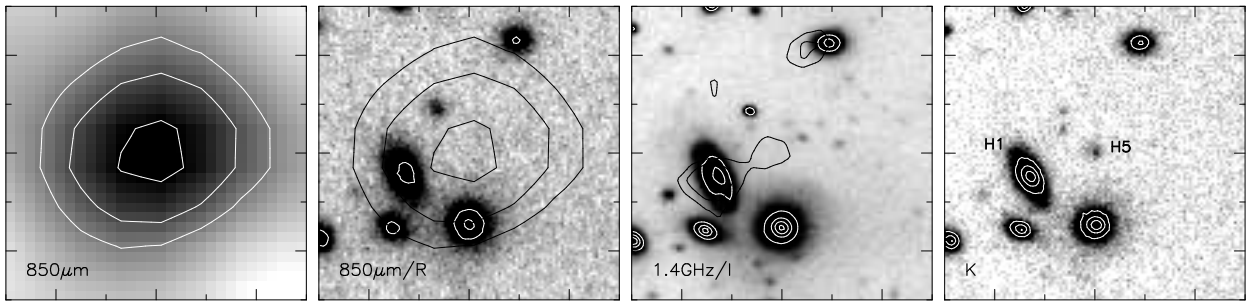}}
\caption{Four different views of the field of SMM\,J09429+4658 from
the optical and near-IR, through the submm to the radio.  These
four panels show the 850-$\mu$m map; the original Hale 5\,m Gunn-$r$
image used by S98 with the 850-$\mu$m map overlayed; the 0.6$''$
resolution, deep Keck\,II $I$-band image of the same field with the
VLA 1.4-GHz map overlayed (the faintest sources visible in the
$I$-band exposure have $I\sim 25.5$); the UKIRT $K$-band image with
the original candidate counterpart, H1, and new ERO candidate, H5,
both marked.  Each panel is 30$''$ square and is centred on the
nominal position of the 850-$\mu$m peak (absolute accuracy of $\ls
3''$), with north top and east to the left.  The relative radio-optical
astrometry for objects in the field is better than $0.4''$ and hence the
radio source close to the bright galaxy at the top of the frame is
not coincident with that galaxy.
}
\end{figure*}

%
%
\begin{figure*}
\centerline{\psfig{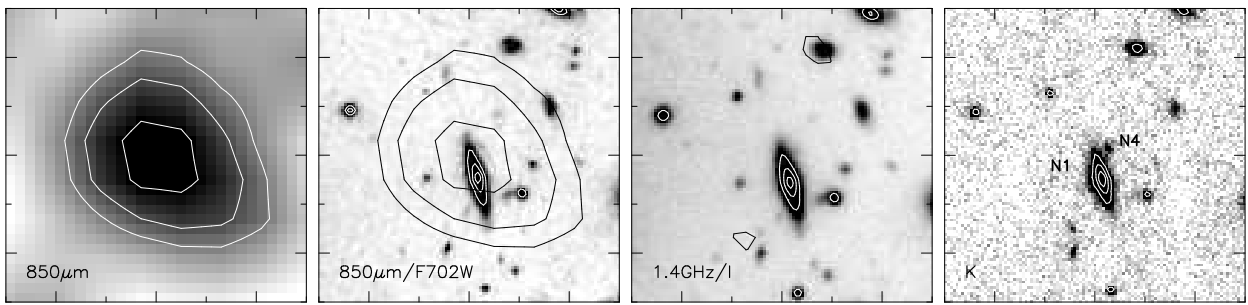}}
\caption{The similar four views shown in Fig.~1, but here for the field
of SMM\,J04431+0210.  The four panels show from left to right, the
850-$\mu$m map; the original {\it HST} F702W identification image used
by S98 with the 850-$\mu$m map overlayed; the deep Keck\,II $I$-band
image of the same field with the VLA 1.4-GHz map overlayed (the
faintest sources visible in the $I$-band exposure have $I\sim 26$); the
UKIRT $K$-band image with the original candidate counterpart, N1, and
new ERO candidate, N4, both marked.   Faint emission coincident with N4
is just visible in the $I$/F702W images, however, aperture photometry
does not confirm this as a formal detection and hence we have instead
quoted a 2$\sigma$ upper limit for N4 in these passbands. Each panel is
30$''$ square and is centred on the nominal position of the 850-$\mu$m
peak (absolute accuracy of $\ls 3''$), with north top and east to the
left.  The relative radio-optical astrometry in the field is better
than $0.4''$.
}
\end{figure*}

\section{Observations}

The two submm sources discussed here were discovered in deep 850-$\mu$m
maps taken with the SCUBA bolometer array on the JCMT\footnote{The
James Clerk Maxwell Telescope is operated by the Joint Astronomy Centre
on behalf of the Particle Physics and Astronomy Research Council of the
United Kingdom, the Netherlands Organisation for Scientific Research,
and the National Research Council of Canada.} during the Lens Survey of
S98.  They are SMM\,J09429+4658 in the field of the cluster
Cl\,0939+4713 (A851) and SMM\,J04431+0210 in the MS\,0440+02 field.
The log of the observations are given in Table~1.  The maps of
SMM\,J09429+4658 were supplemented with data obtained using SCUBA's
photometry mode during the night of 1999 Feb 17. Details of the data
reduction can be found in Ivison et al.\ (1998a, 1998b).

Analysis of deep optical imaging data for the fields indicated that
the most probable counterparts (on the basis of optical magnitudes and
relative positions) for the two sources were both bright spiral galaxies,
H1 and N1 (Fig.~1 and 2), 4.5 and 2.3 arcsec from the SMM\,J09429+4658
and SMM\,J04431+0210 respectively (S98).  H1 has a prominent dust lane
in {\it HST} {\it WFPC2} and {\it NICMOS} images (see also Smail et
al.\ 1999).  Spectroscopy of these galaxies identified N1 as a $z=0.18$
cluster member with relatively strong [O{\sc ii}]\,$\lambda 3727$ and
H\,$\alpha$ emission (Barger et al.\ 1999b), while the redshift of H1
is $z=0.33$ which places it foreground to the cluster in that field.
H1 shows no strong emission lines in its relatively low signal-to-noise
spectrum (Dressler et al.\ 1999).

Until the next generation of millimeter interferometers becomes
available the combination of radio and near-IR data is the cleanest
route to identify reliable counterparts for the submm sources.
Therefore as part of the identification procedure for the S98 submm
survey, we have obtained sensitive 1.4-GHz maps from the VLA and deep
near-IR imaging of all of the sources (Ivison et al.\ in prep).  The
radio maps are sensitive to the same starburst (or AGN activity in any
radio-loud cases) which is powering the submm emission, while at the
same time providing substantially higher astrometric precision and
resolution -- both crucial for correctly identifying the counterparts
to the submm sources.   Combined with the submm fluxes the radio maps
can also provide crude redshift information on these galaxies (Carilli
\& Yun 1999).  In addition, near-IR imaging, described later, offers an
opportunity to identify the dusty, luminous submm galaxies through
their unusual colours.

The VLA observations of MS0440+02 and Cl0939+4713 were obtained in
the A and B configurations, respectively, at 1.4\,GHz. Relevant details
are listed in Table~1. The maps were cleaned and analysed using {\sc
aips}. Details of this complex procedure are given by Ivison et
al.\ (in prep).  The Cl\,0939+4713 map reaches a 1$\sigma$ noise level of
9\,$\mu$Jy\,beam$^{-1}$. The noise level in the MS\,0440+02 map varies
between 15 and 25\,$\mu$Jy\,beam$^{-1}$ due to a strong confusing
source.  The relative astrometry of the radio maps and optical images
are tied together using a large number of radio sources identified with
bright cluster and field galaxies (see for example Smail et
al.\ 1999).  We estimate that the relative positions of sources in the
maps are better than $\ls 0.4''$.

Two faint 1.4-GHz sources are detected in the Cl\,0939+4713 map within
the error box of the submm emission (Fig.~1).  One of these is
associated with the apparently passive but dusty $z=0.33$ spiral, H1
($S_{1.4}=57\mu$Jy, see Smail et al.\ 1999).  The other is only 0.7$''$
from the nominal submm position but had no obvious optical counterpart
in the ground-based imaging available to S98 and the submm source was
therefore provisionally identified with the spiral galaxy pending
further observations (S98; Barger et al.\ 1999b).  A large number of
the brighter galaxies in the Cl\,0939+4713 field are detected in the
VLA map by virtue of its high sensitivity (Smail et al.\ 1999; Morrison
et al.\ 1999).  The radio emission from the spiral population is
interpreted as showing that massive stars are being formed in these
galaxies.  However, a large fraction of these galaxies have no [O{\sc
ii}]$\lambda$3727 emission in their spectra which suggests that this
star formation is highly obscured (Smail et al.\ 1999).     

The lack of radio emission within the submm error box of
SMM\,J04431+0210 in our sensitive VLA map is not particularly unusual.
Roughly half of the submm sources in S98 are undetected in radio
maps at flux limits of $S_{1.4}\sim 0.1$\,mJy (Smail et al.\ 2000).
We place a $3\sigma$ limit of $<70$\,$\mu$Jy on the 1.4-GHz flux of N1.

The other aspect of the identification campaign is the acquisition of
deep near-IR images of the source fields.  These have all been obtained
with the IRCAM3 or UFTI near-IR cameras on the 3.8-m
UKIRT\footnote{UKIRT is operated by the Joint Astronomy Centre on
behalf of the Particle Physics and Astronomy Research Council of the
United Kingdom.} on Mauna Kea (see Table~1).  Observations consist of
deep $K$-band exposures with additional $J$- and $H$-band observations
of the brighter sources detected in $K$.  We aimed to reach
$K\sim20$--21 in our imaging where this limit is set by the need to
identify any counterparts within the submm error-box as unusual from
their extreme optical-near-IR colors, $(I-K)\gs 5$, and the depth of
our available $I$-band images ($I_{\rm lim}\sim 26$ in our multi-orbit
{\it HST} exposures).  Integrating any fainter in $K$ might produce
additional candidate counterparts but we would be unable to identify
these as unusual on the basis of their colours alone.

We have imaged 85\% of the submm sample of S98 so far to a median
3$\sigma$ depth of $K\sim 20.5$.  To this depth, the majority of the
fields show no sources not already seen in the original optical
imaging used for the source identification (S98).  However, in two of
the fields -- those for the sources SMM\,J09429+4658 and
SMM\,J04431+0210 -- we find previously unidentified faint near-IR
objects close to the nominal submm positions (Figs.~1 and 2, Table~2).
Using the naming scheme of S98 these candidate counterparts are called
H5 for SMM\,J09429+4658 and N4 for SMM\,J04431+0210.  H5 is coincident
with the faint radio source detected in the deep VLA map of this
cluster, the radio properties of N4 and H5 are presented in Table~2.

The $K$-band image of SMM\,J04431+0210 was taken with IRCAM3 in
photometric and good-seeing conditions on the night of 1998 September
10. The original observations of SMM\,J09429+4658 were obtained in
non-photometric conditions on 1998 October 19 using the new 1024$^2$
InSb imager UFTI (Leggett 1999), along with further $K$-band
observations of SMM\,J04431+0210.  Each of these exposures consists of a
total of 3.2\,ks of on-source integration, with the UFTI images 
being slightly shallower due to the poor transparency and seeing
encountered.  Nevertheless, the UFTI exposure of SMM\,J04431+0210
provides an independent confirmation of the reality of the source N4.
To check the reliability of the detection of H5 and to improve our
photometry we repeated the observations of SMM\,J09429+4658 using
IRCAM3 in much better conditions on the nights of 1999 February
10--11, obtaining a total of 8.1\,ks integration in $K$ and 3.2\,ks in
$H$. These observations confirm that H5 is indeed an extremely red
source, with only a marginal detection of the source in the $H$-band
image giving $(H-K)\gs 2.4$.

All of the near-IR observations were broken into subsets of nine
exposures dithered on a $3\times 3$ grid. Each of these sets of
sub-exposures was also spatially offset.  The dithering allows us to
construct sky frames from the science exposures and use these to
correct the science frames.  The science and standard star frames were
linearised, reduced and calibrated onto the Cousin's $I$-band
in a standard manner.  The colour term results in an uncertainty
in the final $I$-band photometry of $+0.15$ for sources as red as 
$(R-I)=3$.

After modelling and removing contaminating light from nearby galaxies,
the magnitudes of the two sources were measured in 3.0$''$-diameter
apertures.  These were converted to total magnitudes using aperture
corrections measured from bright compact objects in each field (Table~2, Fig.~3).  The uncertainties quoted on this photometry includes a
contribution due to removal of the contaminating light.

Optical imaging of the two fields was obtained with the LRIS imaging
spectrograph (Oke et al.\ 1995) on the 10-m Keck\,II telescope, Mauna
Kea\footnote{The W.\,M.\ Keck Observatory is a scientific partnership
between the University of California, the California Institute of
Technology and the National Aeronautics and Space Administration, and
was made possible by the generous financial support of the W.\ M.\ Keck
Foundation.}. These images were taken to constrain the $(I-K)$ colours
of the two near-IR sources we identified from the UKIRT imaging.  The
observations were obtained in good conditions during bright time
(Table~1) and consist of multiple 200-s exposures of the two fields
dithered on a non-redundant grid.  The science frames were processed in
a standard manner and flatfielded with twilight flat fields.  To remove
fringing from the frames, subsets of the science exposures were stacked
to construct fringe frames which were then subtracted from the
individual science frames.  This procedure worked well and the final
frames reach 2$\sigma$ sensitivities of $I\sim 25.5$--26 within
3.0$''$-diameter apertures.  Notwithstanding the depth of these images,
neither of the near-IR sources N4 or H5 is reliably detected.  Faint
emission is just visible in the $I$, and perhaps F702W, images of N4
coincident with the near-IR source, our aperture photometry does not
confirm this as a formal detection and so instead we have chosen to
quote a conservative limit.  We therefore give the 2$\sigma$ limiting
$I$-band magnitudes for both counterparts in Table~2 and Fig.~3, along
with the equivalent limits from the original $R$-band data of S98.

%
%
\begin{figure*}
\centerline{\psfig{file=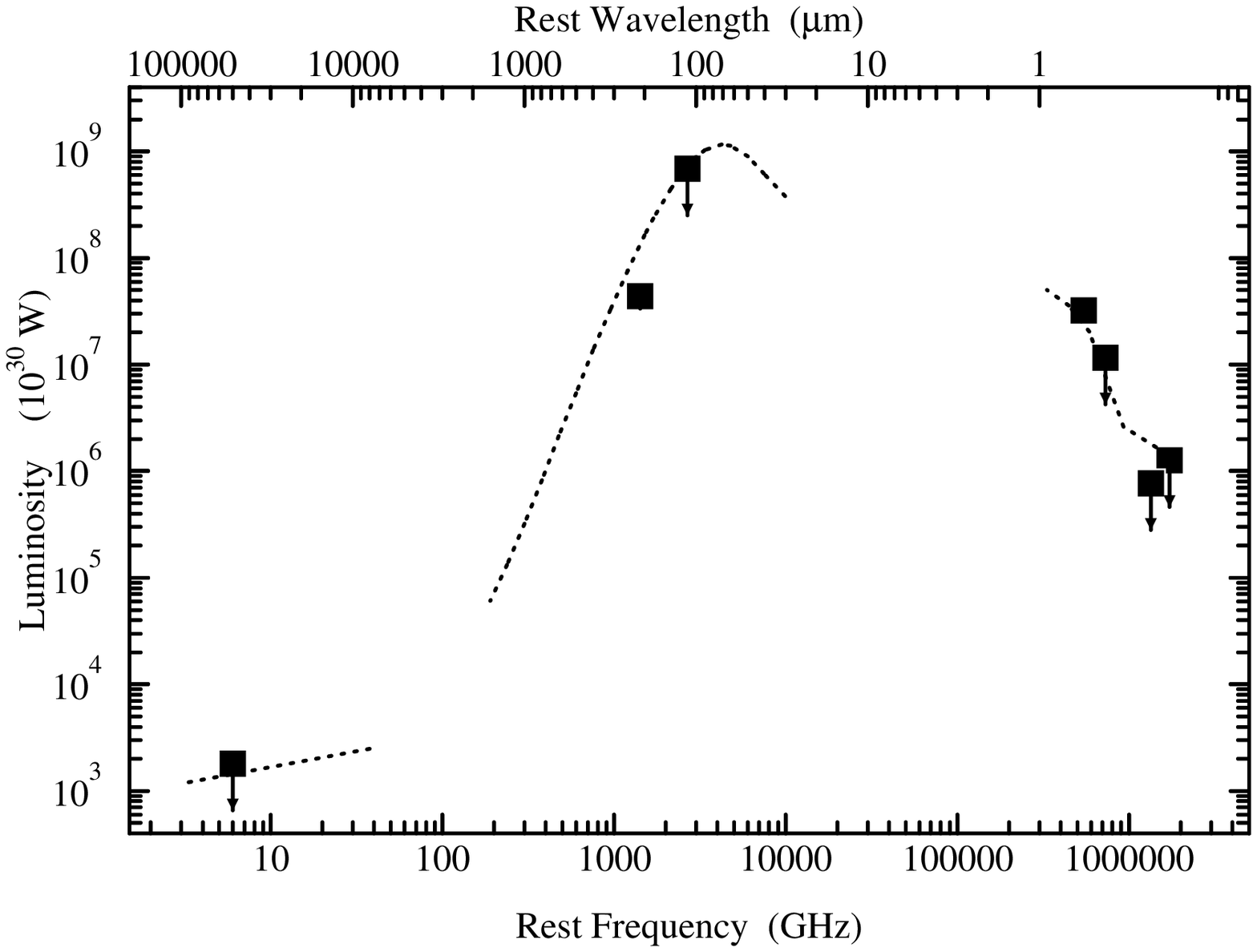,angle=0,width=3in} \hspace{0.4in}
\psfig{file=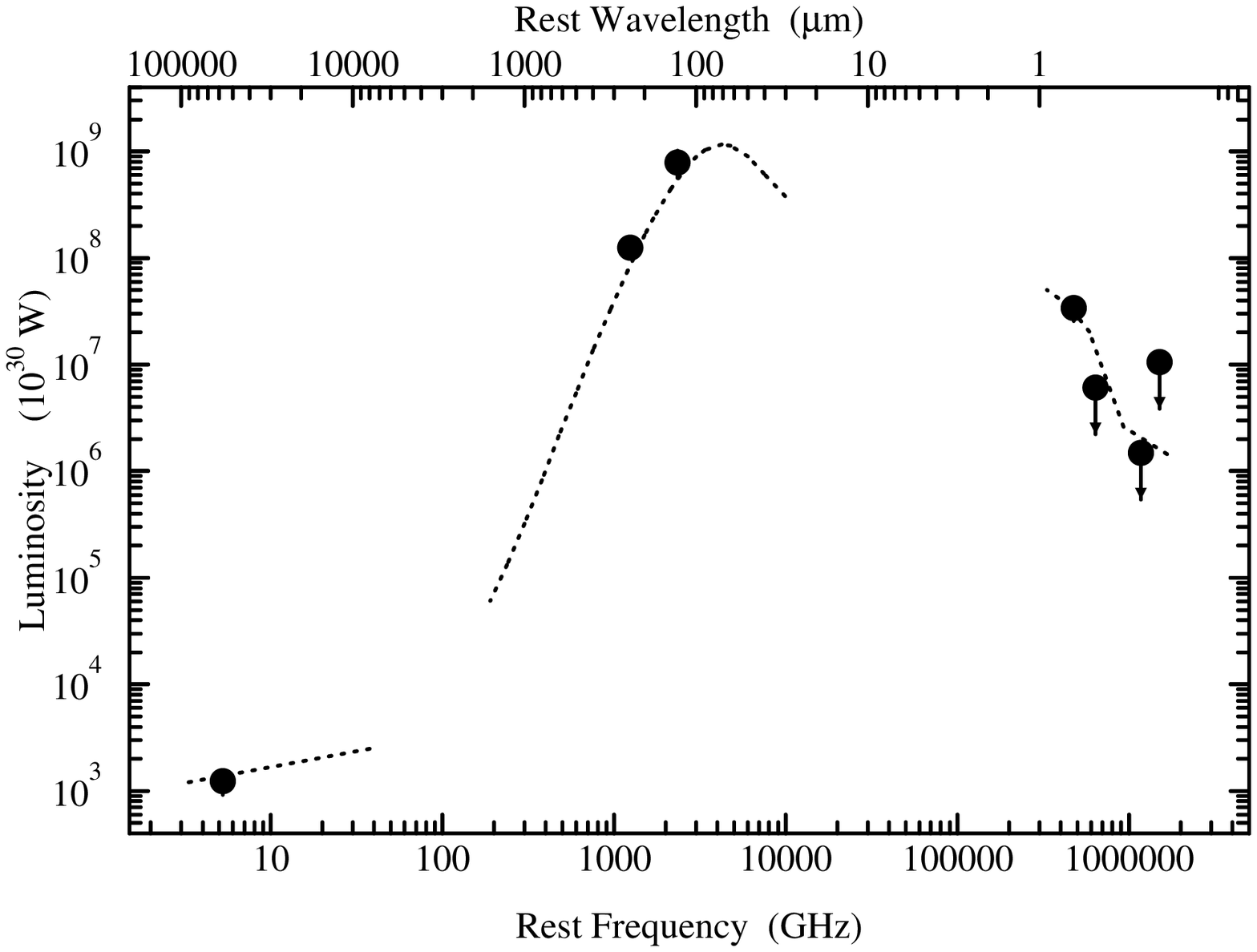,angle=0,width=3in}}
\caption{The rest-frame spectral energy distributions (SEDs) for N4
(left) and H5 (right), corrected for gravitational
amplification, compared to parameterised SEDs for the $z=1.44$
obscured starburst, HR\,10 (dotted lines), based on the available
observations (Dey et al.\ 1999). Redshifts of $z\sim 3$ for N4 and
$z\sim 2.5$ for H5 yield reasonable fits to the data, although the 
allowed range in both cases is large, $\pm 1$.
}
\end{figure*}

\section{Discussion}

Our deep $K$-band imaging of the submm sources from S98 has turned up
two objects with relatively bright near-IR emission which were
invisible on the deep optical imaging used by S98.  We estimate
2\,$\sigma$ limits of $(I-K)\gs 6.8$ and $(I-K)\gs 6.0$ for N4 and H5
respectively.  This puts both N4 and H5 firmly into the class of EROs,
with N4 possibly an extreme example.  Moreover, both objects exhibit
sufficiently red optical to near-IR colours that the probability of
randomly detecting such a source within the submm error boxes is
slight, $P\ll 0.01$. We now discuss the properties of the submm sources
in more detail and argue that the EROs, rather than the bright spirals
N1 and H1, are the most likely counterparts of the submm sources.   

We begin by discussing the spectral indices for the two submm sources
between the submm and radio regimes using our 850-$\mu$m and 1.4-GHz
flux densities or limits.  Carilli \& Yun (1999) have shown that the
spectral index, $\alpha^{850}_{1.4} = 0.42 \times \log_{10}
(S_{850\mu{\rm m}}/S_{1.4{\rm GHz}})$, can be used to obtain a crude
redshift estimate for a range of starburst- and AGN-dominated SEDs.  We
find $\alpha^{850}_{1.4} = 1.02\pm0.07$ for SMM\,J09429+4658 using the
radio flux for H1 or  $\alpha^{850}_{1.4} = 1.10\pm0.08$ using H5's
radio flux and $\alpha^{850}_{1.4} \gs 0.84$ for SMM\,J04431+0210.
These values are at the upper end of the distribution for high-redshift
sources discussed in Carilli \& Yun (1999), with SMM\,J09429+4658
having the highest $\alpha^{850}_{1.4}$ index of any source currently
known. 

Using the various empirical and model curves in Carilli \& Yun (1999),
we translate the $\alpha^{850}_{1.4}$ indices into redshift limits of
$z\gs 4$ for SMM\,J09429+4658 and $z\gs 2$ for SMM\,J04431+0210.  This
indicates that both submm sources lie at high redshifts and are not
associated with the bright spiral galaxies H1 or N1.  The expected
submm--radio spectral index for $z=0.2$--0.3 star-forming galaxies such
as H1 or N1 would be in the range $\alpha^{850}_{1.4} = -0.2$ to 0.3,
far lower than the observed values.  The predicted
submm fluxes for these two galaxies adopting the observed radio
limits/detections and the range of $\alpha^{850}_{1.4}$ for
$z=0.2$--0.3 galaxies from Carilli \& Yun (1999) would be $S_{850}\ls 0.3$\,mJy,
far fainter than the observed fluxes.  The high values of
$\alpha^{850}_{1.4}$ for both SMM\,J09429+4658 and SMM\,J04431+0210
might also suggest that AGN-based emission is unlikely to dominate the
radio power of these sources (Carilli \& Yun 1999).

Another independent redshift estimate for the submm sources is
available to us: the ratio of 450- to 850-$\mu$m flux densities. For
SMM\,J09429+4658, $S_{450}/S_{850} = 4.0 \pm 1.0$, which indicates
$1.5\ls z\ls 5.0$ according to Hughes et al.\ (1998). For
SMM\,J04431+0210, $S_{450}/S_{850} \ls 8.3$, which suggests that
$z\gs0.3$ -- a weak but useful constraint.  The expected ratio of 450-
to 850-$\mu$m flux densities for star-forming galaxies at $z=0.2$--0.3
is $S_{450}/S_{850} \gs 8$.  Again, the spectral properties of the
submm sources appear to be incompatible with the bright galaxies H1 and
N1 being the counterparts of the submm emission and suggest instead
that the source lie at substantially higher redshifts.

Data at submm and radio wavelengths thus appears to rule out the
identification of the two spiral galaxies, H1 and N1, as counterparts
of the submm sources (S98). Therefore we now discuss the properties of the
EROs H5 and N4 to determine if these could be the true counterparts to
the submm sources.

The modest resolution and signal-to-noise of our $K$-band imaging means
we can make no strong statements about the morphologies of the
near-infrared sources H5 and H4.  However, both show signs of being
extended in our 0.5$''$-seeing UKIRT images indicating that they are
likely to be a galaxies. 

As we have shown both submm sources probably lie at high redshifts,
they therefore have bolometric luminosities of $\gs 10^{12} L_\odot$
and their counterparts will class as ULIRGs.  We can thus compare our
limits on the colours of the two EROs with the predicted $(I-K)$
colours of ULIRGs at high redshift from Trentham, Kormendy \& Sanders
(1999).  They use ultraviolet observations of low-redshift ULIRGs from
{\it HST} to model the expected colours of similar systems at high
redshift.  The only galaxy whose colours exceed $(I-K)= 6.0$ (our limit
for H5) is VII\,Zw031, the reddest galaxy in their sample.  This would
have $(I-K)\gs 6.0$ at $z\gs 2$ and $(I-K)\gs 6.8$ at $z\gs 2.5$.  Less
extreme galaxies only become as red as $(I-K)\sim 5.3$ at $z\gs 3$ but
tend to become bluer again after that.  With few observations to
compare with (and a wide range in the possible galaxy colours) this
comparison is necessarily of only limited scope.  Nevertheless, the
colours of N4 and H5 would suggest that N4 may be a highly-obscured
galaxy at $z\gs 2.5$, while H5 could be a similarly obscured galaxy at
$z\gs 2$.

The various constraints on the redshifts of the submm sources and the
possible ERO counterparts are summarised in Table~3.  In each case the
constraints provide a consistent picture suggesting that both the submm
sources and the EROs lie at high redshifts, $z\gs 2$.  Taken together
with the low likelihood of a chance spatial coincidence between a submm
source and an ERO (neither class being particularly numerous), and the
highly obscured nature suggested by the ERO's extreme colours, we
propose that the EROs H5 and N4 are the most likely counterparts of the
submm sources SMM\,J09429+4658 and SMM\,J04431+0210.

We now discuss the consequences of the identification of ERO
counterparts to the submm sources for our understanding of the nature
of these two galaxies, as well as for ERO and submm galaxies in
general.  To simplify the discussion below we adopt redshifts of $z\sim
4$ for H5 and $z\sim 3$ for N4 which are representative of the results
in Table~3.

Before investigating the intrinsic properties of the two galaxies H5
and N2, we must first estimate their lens amplifications. We use our
robust mass models for both cluster lenses (Blain et al.\ 1999b) and
include mass components for not only the main cluster potential but
also the galaxies near the ERO positions using the scaling relations of
Natarajan et al.\ (1998) to estimate their relative contributions.  We
assume redshifts of $z\sim 4$ for H5 and $z\sim 3$ for N4 and estimate
an amplification factor of 2.0 for H5, with a range of 1.5--2.1 for
$z=1$--5; N4 has an amplification of 4.4 at $z\sim 3$, with a range of
3.1--4.8 across $z=1$--5.  These magnifications
are achromatic and have typical errors of around 20\%, comparable to the absolute calibration errors
of the SCUBA maps.  Taking the appropriate amplifications we correct
the observed 850-$\mu$m flux densities and estimate intrinsic apparent
fluxes of $7.7\pm 1.0$\,mJy for H5 and $1.6\pm 0.4$\,mJy for N4 (where
the errors do not include any systematic components due to the unknown
redshifts of the galaxies).  The corrected $K$-magnitudes are 20.1 for
H5 and 20.7 for N4, while the equivalent $I$-band limits for these
galaxies if they had been found in a blank field survey would be very
faint $I\gs 26.3$ and $I\gs 27.6$.

Assuming that the far-IR spectral energy distributions (Fig.~3) of the
two EROs are roughly similar to that of HR\,10 (Dey et al.\ 1999), then
their intrinsic submm fluxes, assuming $T_{\rm dust} = 47$K, correspond
to bolometric luminosities of $19\times 10^{12} {\rm L}_\odot$ and
$5\times 10^{12} {\rm L}_\odot$ for H5 and N4 at $z\sim 4$ and $z\sim
3$ respectively (or a range of 35--$18\times 10^{12}$ and 8--$3\times
10^{12}$ for $z=1$--5).  Thus both galaxies class as ULIRGs.  If we
assume that the far-IR emission from these galaxies is purely due to
star formation, the star-formation rates we derive are of the order of
1000\,M$_\odot$ yr$^{-1}$ for stars above 10\,$M_\odot$ for N4 and
4000\,M$_\odot$ yr$^{-1}$ for H5 (assuming SFR$(M\geq 10M_\odot$
yr$^{-1}) = L_{bol}/(0.5 \times 10^{10} L_\odot)$ see Ivison et
al.\ 1998a; Thronson \& Telesco 1986).  However, in the local Universe,
the majority of galaxies as luminous as H5 show some signs of AGN
activity (Sanders \& Mirabel 1996) and hence it is probable that this
galaxy is a composite starburst/AGN, an obscured system akin to
SMM\,J02399$-$0136 (Ivison et al.\ 1998a).

We next compare the characteristics of H5 and N4 with those of HR\,10,
the submm-bright ERO at $z=1.44$.  HR\,10 has an $(I-K)$ colour of 5.8,
with $K=18.4$, a flux density of 4.9\,mJy at 850\,$\mu$m and a
bolometric luminosity of $7\times 10^{12} L_\odot$ (Dey et al.\ 1999;
c.f.\ Cimatti et al.\ 1998).  Thus HR\,10 has a comparable bolometric
luminosity to N4 and is $\sim 3$ times fainter than H5; in contrast,
HR\,10 is nearly an order of magnitude brighter than either N4 or H5
in the $K$-band.  Assuming similar rest-frame optical/far-IR ratios
for all three galaxies, their relative $K$-band magnitudes are
compatible with N4 and H5 lying at higher redshifts than HR\,10, as
suggested by their $\alpha^{850}_{1.4}$ indices (HR\,10 has
$\alpha^{850}_{1.4} \geq 0.5$, which is consistent with the models of
Carilli \& Yun (1999) for $z=1.44$) as well as their 450- to
850-$\mu$m flux density ratios.  We plot the multi-wavelength SED of
HR\,10 along with the available observations of H5 and N4 in Fig.~3.

The similarly extreme optical/near-IR colours and luminosities for N4,
H5 and HR\,10, as shown in Fig.~3, supports the suggestion that N4 and
H5 are more distant analogs of HR\,10.  The identification of ERO
counterparts to two of the submm sources from S98 indicates that at
least 10\% of the faint submm population down to 850-$\mu$m flux levels
of a few mJy could be EROs.  This estimate would rise to 20\% if the
optical blank-field sources in S98 turn out to be faint EROs, which is
entirely plausible.  Taking the 10\% estimate and the surface density
of 850-$\mu$m sources detected in the SCUBA Lens Survey (Blain et
al.\ 1999b) of 2.5--$4 \times 10^3$ per square degree brighter than
2\,mJy, we would expect a surface density of submm-selected EROs of
around 0.1 per square arcmin.  The $K$-band magnitudes of these
galaxies are $K\sim 20.5$ in the absence of lensing and hence they
would also account for all of the reddest objects in the ERO population
at this limit (Cowie et al.\ 1999).

Our conclusion that all of the most extreme ERO population are likely
to be submm emitters agrees well with the results from pointed SCUBA
observations of extreme EROs (see \S1). These programs are also detecting
such sources at flux densities of $S_{850}\gs 2$\,mJy.  We therefore
suggest that the most extreme EROs, $(I-K)\geq 6$, comprise a population
of dusty ultraluminous starbursts in the distant Universe.  While this
population produces only a minor component of the background, $\ls 10$\%
of the far-infrared background (see Blain et al.\ 1999a), the study of
these sources will be an important step in understanding the formation
and evolution of dust within the most luminous and  obscured galaxies
at high redshifts.
 
We stress that the whole ERO population (defined as those objects redder
than $(I-K)\geq 5$) is not dominated by submm-bright sources. This is
not particularly surprising given the wide range of characteristics
which can place a galaxy in that class (see \S1).  However, the extreme
colours of H5 and N4 and the spread of optical to near-IR colours which
they indicate in the submm-selected sample (c.f.\ Ivison et al.\ 1998a)
is more puzzling and suggests a relatively inhomogeneous population with
a wide range of line-of-sight dust obscuration. Similar behaviour is
seen in samples of low-redshift ULIRGs (Trentham, Kormendy \& Sanders
1999) and these are therefore likely to be the best laboratories for
unravelling the cause of these widely varying optical properties.

We also note that relatively bright ERO counterparts ($K\ls 20.5$) such
as those discussed here are not seen for the bulk of the submm sources
in our survey.  Moreover, in at least  two cases we have submm sources
with bright and blue optical counterparts, two galaxies at $z=2.80$ and
$z=2.56$ (Ivison et al.\ 1998a, 1999), and in both instances we have
confirmed the identifications with CO observations at mm wavelengths
(Frayer et al.\ 1998, 1999).   Hence while the SCUBA population
contains some near-IR bright EROs, these do not appear to dominate the
sample.  It is possible that some of the remaining SCUBA sources have
counterparts with ERO-like characteristics which are fainter than
$K\sim 21$ (equivalent to $K\gs 22$ for a blank field survey). Such objects would be exceedingly difficult to identify and
their further study would be almost completely confined to what could be
learnt at millimeter and radio wavelengths.

The similarities between the properties of H5, N4 and HR\,10 along
with the identification of the latter as a massive, dust-enshrouded
starburst galaxy at $z=1.44$ (Dey et al.\ 1999) suggests that these two
galaxies represent similar systems lying at even higher redshifts and
hence earlier times in the Universe. The redshift constraints from our
multi-wavelength observations are summarised in Table~3.  These massive
star-forming galaxies will provide particularly stringent tests of
hierarchical galaxy-formation models (Baugh et al.\ 1998) if it can be
shown that a large fraction of their bolometric luminosity is powered
by star formation.

\section{Conclusions}

\noindent{$\bullet$} In the course of a near-IR survey of counterparts
to faint submm sources we have uncovered two extremely red galaxies, H5
and N4, which were undetected in the deep optical images used
originally to select likely counterparts by Smail et al.\ (1998).
Follow-up deep optical imaging of both fields with Keck\,II puts
2-$\sigma$ limits of $(I-K)\gs 6.0$ and $(I-K)\gs 6.8$ on the colours
of H5 and N4 respectively. Both galaxies therefore class as EROs. 

\noindent{$\bullet$}  Using the submm and radio spectral properties of
the submm sources we argue that the EROs are probably the  source of
the submm emission, not the bright spiral galaxies previously
identified as such by Smail et al.\ (1998).  The identification of two
ERO counterparts to submm sources indicates that at least 10\% of the
submm population down to 850-$\mu$m flux levels of a few mJy could be
EROs.  The equivalent surface density of submm-selected EROs would be
around 0.1 per square arcmin at $K\ls 20.5$. This density would account
for all of the reddest EROs detected in near-infrared surveys at this
depth.

\noindent{$\bullet$} A comparison of the submm and radio emission from
H5 and N4 against the models of Carilli \& Yun (1999) and a study of
their optical and far-infrared SEDs suggests that both galaxies are
likely to be dusty ultraluminous starbursts at high redshifts, probably
at $z\gs 2$. 
  
\section*{Acknowledgements}

We thank Neil Trentham for his thorough referee's
report on this paper as well as for kindly providing the predicted $(I-K)$
colours of high redshift ULIRGs.  We acknowledge useful conversations
with Arjun Dey, James Graham and Katherine Gunn.  We thank Sandy
Leggett for help and support during our UFTI observations.  IRS
acknowledges support from the Royal Society and RJI and AWB from
PPARC.

%
%
\begin{table*}
\begin{center}
\centerline{\sc Table 1}
\vspace{0.1cm}
\centerline{\sc Log of Observations}
\vspace{0.1cm}
\begin{tabular}{lcccccl}
\hline\hline
\noalign{\smallskip}
{Telescope} & {Instrument} & {Date} & {Band} &  {t$_{\rm exp}$} & {Resol.} & {Comments} \cr
& & & & (ks) & ($''$) & \cr
\hline
\noalign{\medskip}
\multispan2{{\bf SMM\,J09429+4658}} \cr
\noalign{\smallskip}
JCMT & SCUBA & 1998 Mar 12--13 &  850\,$\mu$m & 30.1 & 14.5 & mapping mode \cr
JCMT & SCUBA & ... &  450\,$\mu$m & ... & 7.8 & mapping mode, simultaneous with 850\,$\mu$m \cr
JCMT & SCUBA & 1999 Feb 17&  850\,$\mu$m & 1.8 & 14.7 & photometry mode \cr
JCMT & SCUBA & ... &  450\,$\mu$m & ... & 7.8 & photometry mode, simultaneous with 850\,$\mu$m \cr
VLA  &  ...  & 1996 Jan 06-08 &  21.4\,cm    & 60.7 &  4.0 & B configuration \cr
UKIRT & UFTI & 1998 Oct 19  &  $K$  & 3.2 &  1.0 & non-photometric \cr
UKIRT & IRCAM3 & 1999 Feb 10  &  $K$  & 8.1 &  0.75 & photometric \cr
UKIRT & IRCAM3 & 1999 Feb 11  &  $H$  & 3.2 &  0.65 & photometric \cr
Keck\,II & LRIS & 1998 Nov 01 & $I$ & 3.6 & 0.65 & photometric  \cr
P200 & COSMIC & 1993 Nov 06  &  $r$  & 3.0 &  1.2 & photometric \cr
\noalign{\medskip}
\multispan2{{\bf SMM\,J04431+0210}} \cr
\noalign{\smallskip}
JCMT & SCUBA & 1997 Aug--1998 Sep &  850\,$\mu$m & 35.8 & 14.7 & mapping mode \cr
JCMT & SCUBA & ... &  450\,$\mu$m & ... & 7.8 & mapping mode, simultaneous with 850\,$\mu$m \cr
VLA  &  ...  & 1998 Apr 19  &  21.4\,cm    & 28.0 & 1.4 & A configuration \cr
UKIRT & IRCAM3 & 1998 Sep 10 &  $K$ & 3.2 & 0.50 & photometric  \cr
UKIRT & UFTI & 1998 Oct 19 &  $K$ & 3.2 & 1.1 & non-photometric  \cr
UKIRT & IRCAM3 & 1999 Feb 12  &  $H$  & 3.2 &  0.50 & photometric \cr
Keck\,II & LRIS & 1998 Nov 01 & $I$ & 6.5 & 0.60 & photometric \cr
{\it HST} & {\it WFPC2} & 1994 Oct 06  &  F702W  & 6.5 & 0.10 &  \cr
\noalign{\hrule}
\noalign{\smallskip}
\end{tabular}
\end{center}

\end{table*}

%
%
\begin{table*}
\begin{center}
\centerline{\sc Table 2}
\vspace{0.1cm}
\centerline{\sc Properties of the ERO Counterparts}
\vspace{0.3cm}
\begin{tabular}{lccl}
\hline\hline
\noalign{\smallskip}
 {Property} & {H5} & {N4} & {Comments}\cr
\hline
\noalign{\smallskip}
R.A.\ (J2000) & 09 42 54.65 & 04 43 07.10 & near-IR position \cr
Dec.\ (J2000) & +46 58 44.7 & +02 10 25.1 & accurate to $\pm 1''$ \cr
$\delta {\rm (IR)}$    & 0.9 & 1.9 & arcsec, submm to near-IR offset \cr
$\delta {\rm (radio)}$ & 0.7 & ... & arcsec, submm to radio offset \cr
\noalign{\medskip}
850\,$\mu$m& 15.4$\pm$2.0 & 7.2$\pm$1.7 & mJy, weighted mean of map and  photometry$^\dagger$\cr
450\,$\mu$m& 61$\pm$13    & $3\sigma<60$ & mJy, weighted mean of map and photometry$^\dagger$ \cr
20\,cm     & 36$\pm$9     & $3\sigma<70$  & $\mu$Jy\cr
$K_{\rm tot}$           & 19.39$\pm $0.30 & 19.13$\pm $0.15 & \cr
$K_{\rm ap}$           & 19.58$\pm $0.16 & 19.28$\pm $0.10 & 3.0$''$ diameter aperture photometry\cr
$H_{\rm ap}$           & $>22.0$ & $>21.5$ & 2\,$\sigma$ limits\cr
$I_{\rm ap}$           & $>25.6$ & $>26.0$ & 2\,$\sigma$ limits\cr
$R_{\rm ap}$           & $>24.0$ & $>26.0$ & 2\,$\sigma$ limits from P200/{\it HST} data\cr
\noalign{\hrule}
\noalign{\smallskip}
\end{tabular}

\noindent
$^{\dagger}$ Errors include a 10\% contribution from the uncertainty
in the absolute calibration.
\end{center}
\end{table*}

%
%
\begin{table*}
\begin{center}
\centerline{\sc Table 3}
\vspace{0.1cm}
\centerline{\sc Redshift constraints for the EROs}
\vspace{0.3cm}
\begin{tabular}{lcc}
\hline\hline
\noalign{\smallskip}
{Constraint}         & \multispan2{ \hbox{Redshift}} \cr
                     & {H5}         & {N4}           \cr
\hline
\noalign{\smallskip}
$(I-K)$              & $\gs2$       & $\gs 2.5$  \cr
$\alpha^{850}_{1.4}$ & $\gs4$       & $\gs2$ \cr
$S_{450}/S_{850}$    & $1.5$--$5$   & $\gs0.3$ \cr
SED fit              & $\sim 2.5$   & $\sim 3$ \cr
\noalign{\hrule}
\noalign{\smallskip}
\end{tabular}
\end{center}
\end{table*}

\end{document}